\documentclass[onecolumn,manuscript,showpacs,amsmath,amssymb]{revtex4}
\usepackage{txfonts}
\usepackage{pifont}
\usepackage{amssymb}
\usepackage{dcolumn}
\usepackage{amsmath}
\usepackage[dvips]{epsfig}

\makeatletter
\def\btt#1{\texttt{\@backslashchar#1}}%
\DeclareRobustCommand\bblash{\btt{\@backslashchar}}%
\makeatother

\begin{document}
\title{Anharmonic effect on lattice distortion, orbital ordering and
       magnetic properties in Cs$_{2}$AgF$_{4}$}
\author{Da-Yong Liu$^{1,2}$, Feng Lu$^{1,2}$, and Liang-Jian Zou$^{1,
        \footnote{Correspondence author, Electronic mail:
           zou@theory.issp.ac.cn}}$}
\affiliation{ \it $^1$ Key Laboratory of Materials Physics,
Institute of
              Solid State Physics,\\
       Chinese Academy of Sciences, P. O. Box 1129, Hefei 230031, China\\
             \it $^2$Graduate School of the Chinese Academy of Sciences\\}
\date{June 11, 2008}

\begin{abstract}
We develop the cluster self-consistent field method incorporating
both electronic and lattice degrees of freedom to study the origin
of ferromagnetism in Cs$_{2}$AgF$_{4}$. After self-consistently
determining the harmonic and anharmonic Jahn-Teller distortions, we
show that the anharmonic distortion stabilizes the staggered
x$^{2}$-z$^{2}$/y$^{2}$-z$^{2}$ orbital and ferromagnetic ground
state, rather than the antiferromagnetic one. The amplitudes of
lattice distortions, Q$_{2}$ and Q$_{3}$, the magnetic coupling
strengthes, J$_{x,y}$, and the magnetic moment, are in good
agreement with the experimental observation.
%
%
\end{abstract}

\pacs{71.70.Ej,75.10.-b,75.30.Et}
\maketitle

\section{INTRODUCTION}

  Recently, layered perovskite compound Cs$_{2}$AgF$_{4}$
containing the spin-1/2 4d$^{9}$ Ag(II) ion has attracted great
interest, owing to its isostructural to the high-T$_{c}$ 3d$^{9}$
Cu(II) cuprates. Cs$_{2}$AgF$_{4}$ was first refined to the
tetragonal structure with the space group {\it I4/mmm} in 1974 by
Odenthal et al. \cite{ZAAC407-144}. Very recently, however, McLain
{\it et al.} \cite{NatMat5-561} found that the crystal structure is
orthorhombic with the space group {\it Bbcm}. The magnetic
susceptibility and the inelastic neutron scattering experiments
showed that this compound is well described as a two-dimensional
(2D) ferromagnet (FM) with T$_{c}$ $\thicksim$ 15 K
\cite{NatMat5-561}.  Further, McLain {\it et al.} suggested that
Cs$_{2}$AgF$_{4}$ is orbitally ordered at all temperatures of
measurement, and the orbital ordering (OO) is responsible for the FM.
These properties is in sharp contrast to the high-T$_{c}$ parent
La$_{2}$CuO$_{4}$ that are antiferromagnetic (AFM) insulators.
The microscopic origin of the unusual FM and OO in Cs$_{2}$AgF$_{4}$
attracts a lot of attentions.

  In Cs$_{2}$AgF$_{4}$, the basal plane consists of a 2D lattice of
Jahn-Teller (JT) distorted AgF$_{6}$ octahedra with a pattern of
alternating short and long Ag-F bonds. These analogous compounds
with orbital degeneracy, such as Cs$_{2}$AgF$_{4}$ and
K$_{2}$CuF$_{4}$, {\it etc.} turn out to be 2D FM
\cite{NatMat5-561}, while other compounds, such as K$_{2}$NiF$_{4}$
and Rb$_{2}$MnF$_{4}$, {\it etc.} are AFM \cite{MCB,prl93-226402}.
Notice that the former with active JT ions leads to an orthorhombic
structure, while the latter with nondegenerate orbitals only has a
tetragonal structure. Although the difference between these two
structures seems small, {\it i.e.} the main change is the position
of the fluorine atoms in the basal plane, such difference is of the
utmost importance in leading to the distinct properties
\cite{NatMat5-513}. From early studies
\cite{prb26-6438,pr35-187,JPCS31-2209,prb11-3052} in perovskite
compounds, it is known that the anharmonic JT effect is a decisive
factor for the orthorhombic crystal structure. To date, a lot of
studies have been done for Cs$_{2}$AgF$_{4}$ utilizing the density
functional theory
\cite{prb73-214420,prb76-024417,prb76-155115,CM18-3281,prb76-054426,
prl99-247210}, however, few authors focus on the anharmonic JT
effect on the lattice, OO and FM groundstate properties in
Cs$_{2}$AgF$_{4}$.

Previous studies in analogous compound K$_{2}$CuF$_{4}$ have
provided us two distinct scenarios for the FM and OO ground state.
On the one hand, Kugel and Khomskii showed that the cooperative JT
effect, especially the anharmonic JT effect \cite{ssc13-763}, play a
key role in stabilizing the OO ground state in K$_{2}$CuF$_{4}$.
On the other hand, to resolve the difficulty of the Kugel-Khomskii
(KK) electronic superexchange (SE) coupling model \cite{KK}, which
is usually suitable for the Mott-Hubbard insulators with U$_{d}$ $<$
$\Delta$ (here U$_{d}$ is the on-site Coulomb repulsion and $\Delta$
is the charge transfer energy), and to address the FM and OO ground
state, Mostovoy and Khomskii \cite{prl92-167201} proposed a modified
SE coupling model, which is suitable for the charge transfer
insulating K$_{2}$CuF$_{4}$ with U$_{d}$ $>$ $\Delta$.
Obviously, to understand the unusual groundstate properties, such as
the lattice, orbital and magnetic properties in K$_{2}$CuF$_{4}$ and
Cs$_{2}$AgF$_{4}$, one should incorporate the cooperative JT effect
\cite{jap31-14S} and the anharmonic JT effect into the
charge-transfer SE interactions.

To treat these strongly correlated systems more precisely, based on
the cluster self-consistent field approach developed previously
\cite{IJMPB21-691}, we explicitly take into account the orthorhombic JT
distortions and the charge-transfer SE interactions, in
which spin order, OO and lattice distortion are determined
self-consistently. We show that driven by strong anharmonic effect
and Hund's coupling, Cs$_{2}$AgF$_{4}$ has a much more stable
ferromagnetic ground state. The theoretical amplitudes of lattice
distortions, Q$_{2}$ and Q$_{3}$, the magnetic coupling
strengthes, J$_{x,y}$, and the magnetic moment, are in good
agreement with the experimental observation in Cs$_{2}$AgF$_{4}$.
This paper is organized as follows: an effective Hamiltonian and the
cluster self-consistent field (SCF) approach are described in {\it
Sec. II}; then the results and discussions about the lattice
structure and distortions, orbital ordering and magnetic properties in
Cs$_{2}$AgF$_{4}$ are presented in {\it Sec. III}; the last section
is devoted to the remarks and summary.

\section{Model Hamiltonian and Method}

An effective 2D Hamiltonian in Cs$_{2}$AgF$_{4}$ including both the SE
and the JT couplings is written:
\begin{eqnarray}
  H=H_{QJT}+H_{SE}
\end{eqnarray}
where H$_{SE}$ describes the highly symmetrical
effective SE couplings between spins and orbitals; H$_{QJT}$ describes
the JT couplings associated with the electron and the lattice distortion.
Note that spin, orbital and lattice degrees of freedom couple to each
other in Hamiltonian (1).
The JT effect associated with one hole per Ag$^{2+}$ site in the AgF$_{4}$
sheets reads \cite{jap31-14S}
\begin{eqnarray}
   H_{QJT} &=& g\sum_{i}(Q_{i2}\tau_{i}^{x}+ Q_{i3}\tau_{i}^{z})+
   \frac{K}{2}\sum_{i}(Q_{i2}^{2}+Q_{i3}^{2})
      \nonumber\\
           && +G\sum_{i}[(Q_{i3}^{2}-Q_{i2}^{2})\tau_{i}^{z}
        -2Q_{i2}Q_{i3}\tau_{i}^{x}]
\end{eqnarray}
where both the linear and the quadratic vibronic coupling terms have
been included. The first and second terms describe the linear
harmonic JT effect. The third one, {\it i.e.}, the quadratic
coupling, arises from the anharmonic JT effect and contributes to
the anisotropic energy \cite{ssc13-763, jap31-14S}. Here Q$_{i2}$
and Q$_{i3}$ are the normal vibration coordinates, defined as
$Q_{2}=(-X_{1}+X_{2}+Y_{3}-Y_{4})/2$ and
$Q_{3}=(-X_{1}+X_{2}-Y_{3}+Y_{4}+2Z_{5}-2Z_{6})/\sqrt {12}$
\cite{JCP7-72} with X, Y and Z being the coordinates of the $i-th$ F
ions. And $g$ is the linear JT coupling strength, $G$ is the
coefficient of quadratic coupling, and $K$ is the elastic constant.
For Cs$_{2}$AgF$_{4}$, we fix $K$ $=$ 10 eV/$\AA^{2}$ throughout
this paper.

Since Cs$_{2}$AgF$_{4}$ is strongly correlated charge-transfer
insulator \cite{prl99-247210}, the 2D SE interactions between
4d orbitals of Ag$^{2+}$ ions through 2p orbitals of $F$ atoms
are described \cite{prl92-167201}:
\begin{eqnarray}
  H_{SE} = &&\sum_{\substack{i, \alpha\\\alpha=x,y}}
  [(J_{1}+J_{2}I_{i}^\alpha+J_{3}I_{i}^\alpha I_{i+\alpha}^\alpha)\vec{S}_{i}
            \cdot\vec{S}_{i+\alpha}
\nonumber\\
       && +J_{4}I_{i}^{\alpha}+J_{5}I_{i}^\alpha I_{i+\alpha}^\alpha
       ],
\end{eqnarray}
here the operator $\vec{S}_{i}$ denotes the S $=$ 1/2 spin at site
{\it i}, and $I_{i}^\alpha=\cos\left(2\pi
m_{\alpha}/3\right)\tau_{i}^z-\sin\left(2\pi m_{\alpha}
/3\right)\tau_{i}^x$, which is the combination of the components of
the orbital operators $\vec{\tau}$. $\alpha= x, y$, and ($m_{x}$,
$m_{y}$)=($1$, $2$), denote the direction of a bond in the AgF$_{4}$
sheets. In Eq. (3), the coupling coefficients J$_n$ (n=1-5), read:
\( J_{1}=t^2\left[1/U_{d}+2/(2\Delta+U_{p})
 -J_{H}/(2U_{d}^{2})\right]\),
\( J_{2}=4t^2\left[1/U_{d}+2/(2\Delta+U_{p})\right]\),
\( J_{3}=4t^2\left[1/U_{d}+2/(2\Delta+U_{p})
 +J_{H}/(2U_{d}^{2})\right]\),
\( J_{4}=t^2\left[2/\Delta-1/U_{d}-2/(2\Delta+U_{p})\right]\)
and
\( J_{5}=t^2\left[1/U_{d}+2/\Delta-2/(2\Delta+U_{p})
 +3J_{H}/(2U_{d}^{2})\right]\) with t=$t_{pd}^2/\Delta$, respectively.
Here $\Delta$ is the electron transfer energy between silver 4d to
fluorin 2p orbitals; U$_{d}$, U$_{p}$ are the Coulomb repulsion
energies on Ag and $F$, respectively; and J$_{H}$ is the Hund's
coupling on Ag. Note that the coefficients may be positive and
negative, thus the AFM and the FM components coexist in J$_{n}$.
Since the 4d orbitals of Ag$^{2+}$ ions are less localized than the
3d orbitals of Cu$^{2+}$ ions in K$_{2}$CuF$_{4}$, the Coulomb
repulsion U$_{d}$ of Ag$^{2+}$ ion is about 3 $\sim$ 5 eV, and
J$_{H}$ ranges from 0.1 eV to 0.5 eV, in accordance with the LDA+U
results \cite{prb76-155115,prl99-247210,prb76-024417}. In general,
the Coulomb repulsion U$_{p}$ is about 5 eV in F$^{-}$. The charge
transfer energy $\Delta$ is roughly estimated from the difference of
the centers of gravity between the 4d and 2p levels in band
structures, which also allows to estimate $t_{pd}$ from the
bandwidth of electronic structures available
\cite{prb76-155115,prb76-024417,prl93-226402}, $\Delta$=U$_{d}$-2
eV, and $t_{pd}$ $\simeq$ 0.6 eV.

In such a strongly correlated spin-orbital-lattice system, according
to Feynman-Hellman theorem, the ground state energy is minimized
with respect to Q$_{i2}$ and Q$_{i3}$, {\it i.e.},
\begin{eqnarray}
\langle \frac{\partial H}{\partial Q_{i2}} \rangle = 0, \langle
\frac{\partial H}{\partial Q_{i3}} \rangle = 0.
\nonumber
\end{eqnarray}
From which one could find the strength of the normal modes and the
lattice distortion critically depend on the orbital polarization
through the following equations
\begin{eqnarray}
    \langle Q_{i2} \rangle &=& g\frac{K\langle \tau_{i}^{x} \rangle
    +4G\langle \tau_{i}^{x}
    \rangle\langle \tau_{i}^{z} \rangle}{4G^{2}(\langle \tau_{i}^{x}
    \rangle^{2}+\langle \tau_{i}^{z} \rangle^{2})-K^{2}}
   \nonumber\\
   \langle Q_{i3} \rangle &=& g\frac{K\langle \tau_{i}^{z} \rangle
   +2G(\langle \tau_{i}^{x}
  \rangle^{2}-\langle \tau_{i}^{z} \rangle^{2})}{4G^{2}(\langle \tau_{i}^{x}
    \rangle^{2}+\langle \tau_{i}^{z} \rangle^{2})-K^{2}}
\end{eqnarray}
Notice that in the absence of the anharmonic JT effect (G=0),
$<$Q$_{i2}$$>$ $\sim$ $-$(g/K)$<$$\tau_{i}^{x}$$>$ and
$<$Q$_{i3}$$>$ $\sim$ $-$(g/K)$<$$\tau_{i}^{z}$$>$. To obtain the
amplitude of the lattice distortion, one should determine the spin,
the orbital and the deformation configurations self-consistently.

  In order to treat the spin-orbit correlations and fluctuations
with high accuracy in strongly correlated system, the cluster-SCF
approach \cite{IJMPB21-691} developed previously is applied to the
spin-orbital Hamiltonian (1). The cluster-SCF approach includes the
exact treatment to the interactions inside the cluster, and the
self-consistent field treatment on the interactions between the
cluster and the surrounding environment.
 The main procedure is shortly outlined as follows: first,
we choose a cluster consisting of 4 Ag$^{2+}$ ions. The $i-th$ site
is surrounded by two Ag sites inside the cluster and two Ag sites
outside the cluster, which provide two internal interactions and two
external SCF fields, respectively. Thus, the effective Hamiltonian
of the cluster reads:
\begin{eqnarray}
 h_{cluster} & = & \sum_{\substack{i, \alpha\\\alpha=x,y}} [
  (J_{1}+\frac{J_{2}}{2}(I_{i}^\alpha+I_{i+\alpha}^\alpha)
  +J_{3}I_{i}^\alpha I_{i+\alpha}^\alpha)
   \vec{S}_{i}\cdot\vec{S}_{i+\alpha}
  \nonumber \\
 &&+ J_{4}I_{i}^\alpha
 + J_{5}I_{i}^\alpha I_{i+\alpha}^\alpha ]
 + g\sum_{i}(Q_{i3}\tau^{z}_{i}+{Q_{i2}\tau^{x}_{i}})
 \nonumber \\
 &&+ G\sum_{i}[(Q_{i3}^{2}-Q_{i2}^{2})\tau_{i}^{z}
   -2Q_{i2}Q_{i3}\tau_{i}^{x}] \nonumber\\
 && + \frac{K}{2}\sum_{i}(Q_{i2}^{2}+Q_{i3}^{2})+ \sum_{i} h^{scf}_{i}
\end{eqnarray}
with the SCF h$^{scf}_{i}$ contributing from the $j'-th$ external
site interacting with the $i-th$ internal site through H$_{ij'}$,
   $h^{scf}_{i} = Tr_{j'}(\rho_{j'}H_{ij'})$,
where $\rho_{j'}$ denotes the reduced density matrix of the $j'th$
site, and {\it i} runs over all sites inside the cluster.
We first substitute the spin coupling
$\vec{S}_{i}\cdot\vec{S}_{i+\alpha}$ into the cluster Hamiltonian
with the initial spin correlation function
$\langle\vec{S}_{i}\cdot\vec{S}_{i+\alpha}\rangle$, and diagonalize
the orbital part of the cluster Hamiltonian (5) in the presence of
the orbital SCF. The orbitalization $\langle\vec{\tau}\rangle$ and
the orbital correlation functions
$\langle\vec{\tau}_{i}\cdot\vec{\tau}_{i+\alpha}\rangle$ are thus
obtained.
And then, we substitute the orbital operator and the orbital
couplings with $\langle\vec{\tau}\rangle$ and
$\langle\vec{\tau}_{i}\cdot\vec{\tau}_{i+\alpha}\rangle$ into
$h_{cluster}$, and diagonalize the spin part of the cluster
Hamiltonian in the presence of the spin SCF. Hence, we obtain a set
of new averaged spin $\vec{S}_{i}$ and spin correlation functions
$\vec{S}_{i}\cdot\vec{S}_{i+\alpha}$. Repeat the above steps until
the groundstate energy and the spin and orbital correlation
functions converge to the accuracies. From the stable spin-orbital
ground state, one could get the spin coupling strengths J$_{x}$ and
J$_{y}$ through J$_{\alpha}$= J$_{1}$+J$_{2}$$<I_{i}^{\alpha}>$+
J$_{3}$$<I_{i}^{\alpha}I_{i+\alpha}^{\alpha}>$, here $\alpha=x,y$.
The advantage of our approach over the traditional mean-field method
is that the short-range spin and orbital correlations as well as the
quantum fluctuations are taken into account properly, especially in
the low-dimensional systems.

\section{RESULTS AND DISCUSSIONS}

  In this section, we present the numerical results of the ground state
of the charge-transfer insulator Cs$_{2}$AgF$_{4}$ within the
cluster-SCF approach. We mainly discuss the role of the anharmonic
effect on lattice structure and distortions, orbital ordering and magnetic
properties in the ground state.

\subsection{Groundstate Phase Diagram}

  To clarify the roles of the SE coupling and the JT effect on the
origin of the FM, we first perform numerical calculations on Eq. (5)
and obtain the g-J$_H$ phase diagram, as shown in Fig. 1. Since the on-site
Coulomb interaction U$_{d}$ in Cs$_{2}$AgF$_{4}$ is not well defined, we
present the g-J$_H$ phase diagram for U$_{d}$=3, 4, and 5 eV in Fig. 1(a),
1(b) and 1(c), respectively.
Accordingly, we find that the anharmonic distortion strength G plays
a very important role in the phase diagram. With the variation of G
from negative to positive via zero, the FM-AFM phase boundaries
exhibit critical changes. As we see in Fig. 1(a), at G=$-$0.75g, the
stable groundstate phase in the regions I, II, III and IV is AFM,
while that in the region V is FM; at G=0, a vertical line separates
the AFM ground state in the regions I, II, and III from the FM one
in the regions IV and V; when the anharmonic distortion strength G
becomes positive, more regions become FM ordering; at G=0.75g, only
the ground state in the region I is AFM. Fig. 1(b) and Fig. 1(c)
qualitatively resemble to Fig. 1(a).

The competitive FM and AFM couplings in the SE Hamiltonian result in
the FM phase for large J$_{H}$ and the AFM phase for small J$_{H}$
in these phases. As we expect in Fig. 1(a) to Fig. 1(c), strong
Coulomb repulsion favors AFM phase, hence the AFM regime becomes
large with the increase of the U$_d$.
Besides the fact that strong Hund's coupling J$_{H}$ favors the FM
phase, the large anharmonic JT effect also favors the FM ground
state. Therefore, the complete consideration of the realistic
interactions in Eq. (1) most probably leads to the FM ground state
in Cs$_{2}$AgF$_{4}$. And from the recent first-principles
electronic calculations, the Coulomb repulsion between Ag 4d
electrons is about 3 eV \cite{prb76-155115}. We expect that the
Hund's coupling J$_{H}$ lies in 0.1 $\sim$ 0.3 eV, so that the
interaction parameters of Cs$_{2}$AgF$_{4}$ fall in the FM region in
the phase diagram.
%
\begin{figure}[htbp]\centering
\includegraphics[angle=0, width=0.4 \columnwidth]{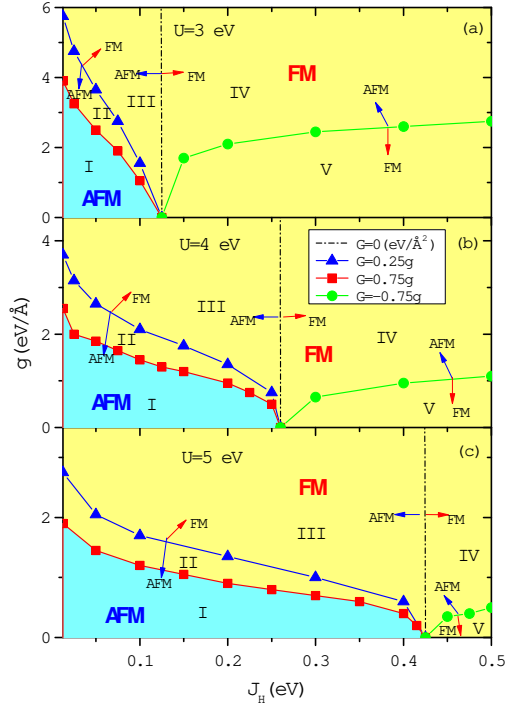}
\caption{(Color online) Phase diagrams of JT coupling g versus
Hund's coupling J$_{H}$ in Cs$_{2}$AgF$_{4}$ for different Coulomb
repulsions (a) U$_d$=3 eV, (b) U$_d$=4 eV, and (c) U$_d$=5 eV. The
solid circles, dashed line, solid triangles and solid squares denote
the FM/AFM phase boundaries for the anharmonic coupling G=-0.75g, 0,
0.25g and 0.75g, respectively. G is in the unit of eV/$\AA^{2}$.}
\label{fig1}
\end{figure}

%
%
   Experimentally, it is hard to determine the sign of the
anharmonic coupling strength G in Cs$_{2}$AgF$_{4}$.
From Fig. 1, it is seen that when positive anharmonic effect is
very large, the ground state stabilizes in the FM phase
even in the absence of Hund's coupling.
Therefore the positive anharmonic coupling most favors the FM
ground state.
On the contrary, for G $<$ 0, due to the SE coupling, the ground
state is FM when the harmonic JT coupling strength g is small;
however, when g becomes large, the anharmonic coupling favors the
AFM ground state. Thus, considering the strong JT effect in
Cs$_{2}$AgF$_{4}$, one expects that G is most probably positive,
which will be further confirmed in the lattice distortion, the
orbital ordering and the magnetic properties in what follows.

\subsection{Lattice Structure and Distortions}
   Different from La$_{2}$CuO$_{4}$, the unusual FM and the orthorhombic
structure of Cs$_{2}$AgF$_{4}$ is the result of the interplays of
the spin, orbital and lattice degrees of freedom, especially the
contribution of the anharmonic JT effect. The normal coordinates
Q$_{2}$ and Q$_{3}$ of the JT distortions in Cs$_{2}$AgF$_{4}$ are
obtained self-consistently through Eq. (4) and (5). We find that for
positive and large G, the structural distortion in the ground state
is antiferro-type, (Q$_{2}$, Q$_{3}$; $-$Q$_{2}$, Q$_{3}$),
corresponding to the alternative short and long bonds in the
AgF$_{4}$ plaquette. However, the groundstate energy of the
ferro-type (Q$_{2}$, Q$_{3}$; Q$_{2}$, Q$_{3}$) distortion is higher
than that of the antiferro-type distortion, as shown in Fig. 2. And
the ferro-type distortion only appears in the strong JT region
because of its instability in this system.
%
\begin{figure}[htbp]\centering
\includegraphics[angle=0, width=0.4 \columnwidth]{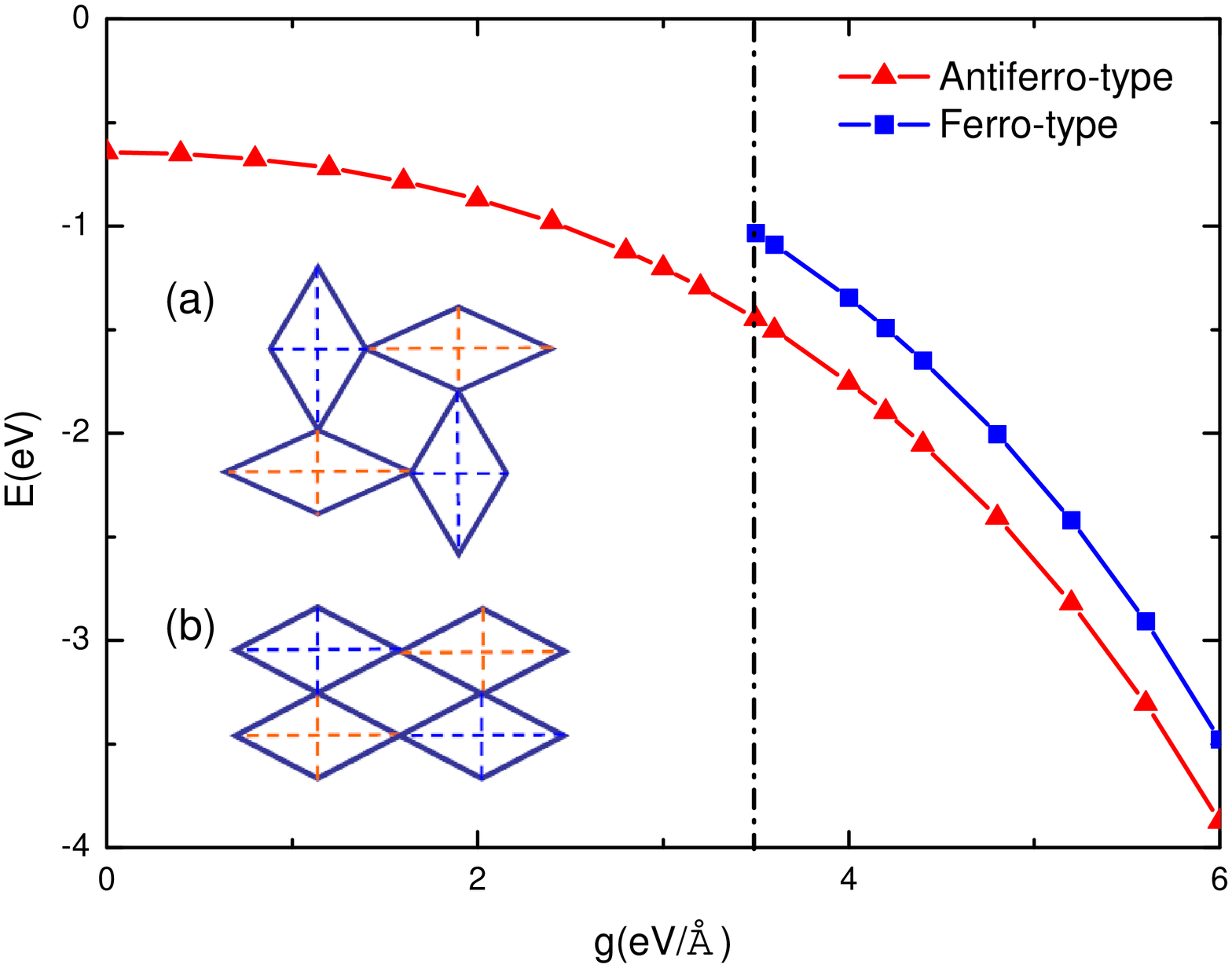}
\caption{(Color online) Dependence of groundstate energy on the JT
coupling strength with G $=$ 0.75g in the antiferro-type (a) and the
ferro-type (b) distortions in the AgF$_4$ plaquette. The dashed
vertical line indicates the appearance of the ferro-type distortion.
Theoretical parameters: U=3 eV and J$_{H}$=0.3 eV.}
\label{fig2}
\end{figure}
%
%
%

  Notice that the experimentally observed lattice structure has the
antiferro-type distortion with the long and short Ag-F bond lengths
of 2.441 and 2.111 $\AA$ in the AgF$_4$ plaquette
\cite{NatMat5-561}, one can deduce the corresponding distortions
Q$_{2}$ $\approx$ 0.33 $\AA$ and Q$_{3}$ $\approx$ $-$0.19 $\AA$.
Theoretically, through the self-consistent numerical calculations,
we obtain the amplitudes of the Q$_{2}$ and the Q$_{3}$ distortions
in Cs$_{2}$AgF$_{4}$ with the variation of the JT coupling for
different anharmonic coupling parameters, as shown in Fig. 3.
When the linear JT coupling strength g is about 4.8 eV/$\AA$ at
G=0.75g, the theoretical distortions are $|Q_{2}|$ $\approx$ 0.33
$\AA$ and Q$_{3}$ $\approx$ $-$0.19 $\AA$, see the dashed vertical
line in Fig. 3, in good agreement with the corresponding
experimental observation in Cs$_{2}$AgF$_{4}$ \cite{NatMat5-561}.
Meanwhile, we find that the positive G leads to the negative Q$_{3}$,
giving rise to the correct distortion c/a $=$ 2s/(l+s) $<$ 1 in orthorhombic
Cs$_{2}$AgF$_{4}$. However, the negative G gives a positive Q$_{3}$, resulting
a wrong distortion c/a $=$ 2l/(l+s) $>$ 1.
%
%
%
%
\begin{figure}[htbp]\centering
\includegraphics[angle=0, width=0.4 \columnwidth]{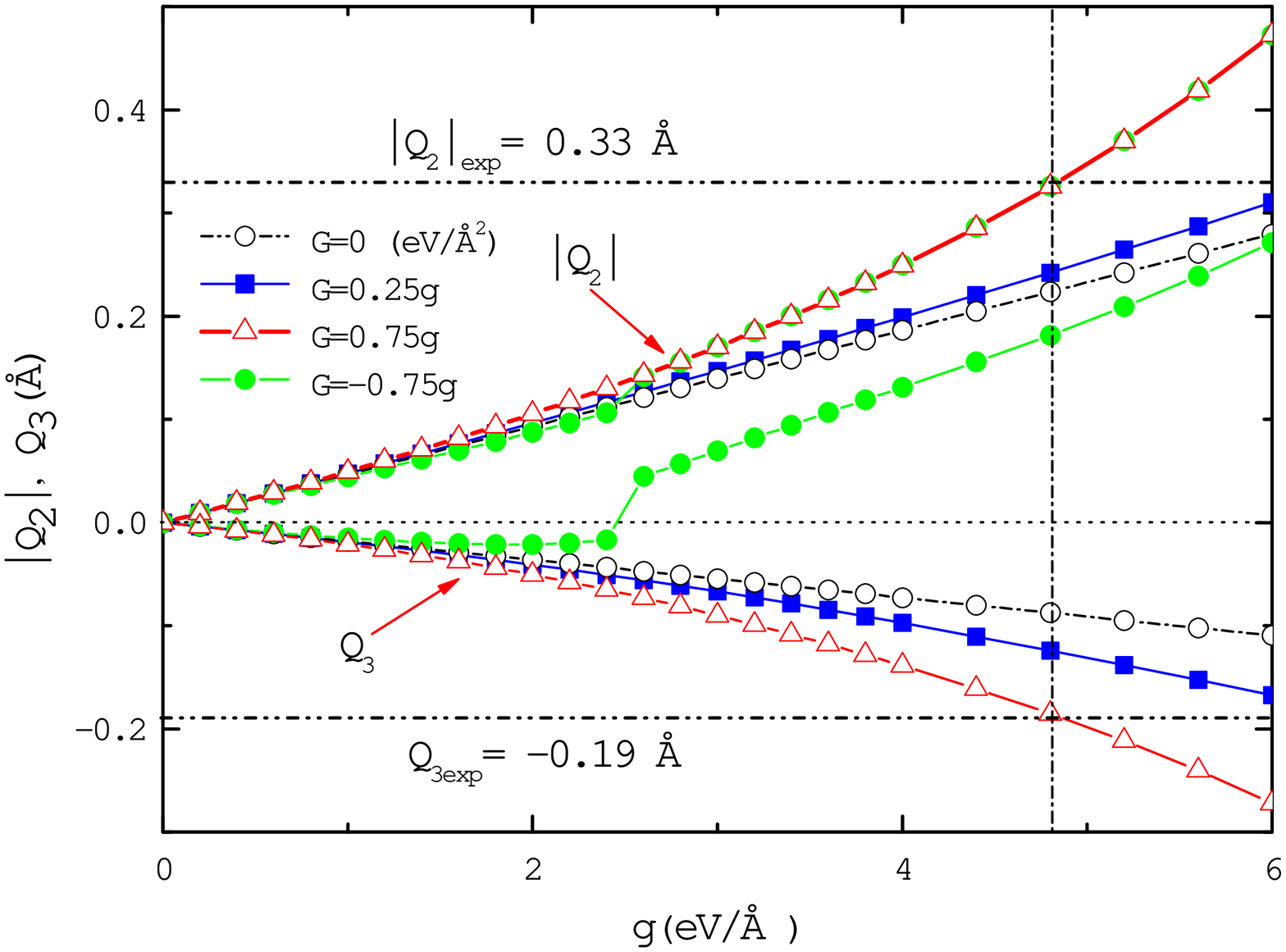}
\caption{(Color online) The amplitudes of $|Q_{2}|$ and Q$_{3}$ as a
function of the JT coupling with different G. Two dashed horizontal
lines correspond to the experimental values of $|Q_{2}|$ and
Q$_{3}$; the vertical dashed line indicates the linear JT coupling
g=4.8 eV/$\AA$ for G=0.75g in Cs$_{2}$AgF$_{4}$. Other parameters
are the same to Fig. 2}
\label{fig3}
\end{figure}
%
%
%
%
All the other analogous substances without JT ions, {\it e.g.}
K$_{2}$NiF$_{4}$, K$_{2}$MnF$_{4}$ \cite{prl93-226402,MCB}, are
tetragonal, rather than the orthorhombic symmetry in
Cs$_{2}$AgF$_{4}$ with JT ions. In fact, it is the positive
anharmonic effect that lowers the lattice symmetry, leading to the
compression of the ligand octahedron along the c-axis \cite{SPU}.

\subsection{Orbital Ordering}

  The unusual FM and distinct orthorhombic structure in Cs$_{2}$AgF$_{4}$
are in fact associated with the formation of the long-range OO, as
shown by a few authors recently utilizing the first-principles
calculations \cite{prl99-247210, prb76-024417, prb76-155115,
prb76-054426}. Our study also confirms this point. More further, we
demonstrate that the anharmonic effect plays an essential role in
the OO properties of Cs$_{2}$AgF$_{4}$.
In general, one can describe the orbital occupied state of each site
in terms of $|\phi \rangle$=cos$\frac{\theta}{2}$ $|3z^2-r^2\rangle$
$\pm$ sin$\frac{\theta}{2}$$|x^2-y^2\rangle$ with the orbital angle
$\theta$. Here $'\pm'$ refers to the two sublattices of the
antiferro-distortion in AgF$_{4}$ plaquette.
Fig. 4 shows the orbital angle $\theta$ as a function of the linear
JT coupling strength g with different anharmonic coupling G. The role of the
anharmonic coupling on the orbital angle, hence the OO, is clearly seen.
For strong JT distortion and G=0.75 g, the OO is staggered {\it
$z^2-x^2$}/{\it $z^2-y^2$}, see inset (b) in Fig. 4. The
orbital angle obtained is about 61.8$^{\circ}$ for g=4.8 eV/$\AA$,
as indicated by the dashed vertical line in Fig. 4,
which is consistent with the experimental suggestion for Cs$_{2}$AgF$_{4}$
\cite{NatMat5-561}.
%
%
This is different from the
case in La$_{2}$CuO$_{4}$, where the distance between apical O and
Cu is much larger than that in the a-b plane, resulting in the pure
$3z^2-r^2$ hole orbitals.
%
%

%
\begin{figure}[htbp]\centering
\includegraphics[angle=0,width=0.4 \columnwidth]{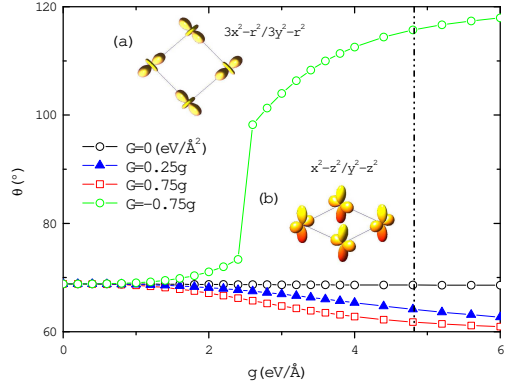}
\caption{Orbital angle as a function of the JT coupling with
different anharmonic coupling. Inset (a) and (b) correspond to the
3$x^{2}-r^{2}$/3$y^{2}-r^{2}$ and $x^{2}-z^{2}$/$y^{2}-z^{2}$
orbital orders for large negative and positive G values,
respectively. The vertical dashed line corresponds to the case of
g=4.8 eV/$\AA$ for Cs$_{2}$AgF$_{4}$. Other parameters are the same
to Fig. 2.}
\label{fig4}
\end{figure}
%
As a comparison, for G $<$ 0, the orbital angle $\theta$ $>$ $\pi/2$;
as $|G|$
becomes very large, $\theta$ $\rightarrow$ $2\pi/3$, and the holes
orderedly occupy the d$_{3x^{2}-r^{2}}$/d$_{3y^{2}-r^{2}}$ orbitals,
as seen the inset (a) in Fig. 4.
%
%
For G $=$ 0, however, the orbital angle is nearly constant with the
increase of the linear JT coupling g, giving rise to
$\theta$=68$^{\circ}$. Note that in the absence of the anharmonic
effect, the SE couplings contribute to the same OO as the harmonic
JT effect does. In this situation, the orbital configuration is
characterized by orbital angle $\theta_{0}$:
$tan\theta_{0}=Q_{i2}/Q_{i3}$ \cite{jap31-14S}. Then $Q_{i2}/Q_{i3}$
$\sim$ $\tau_{i}^{x}/\tau_{i}^{z}$, which is almost independent of
the linear coupling g, indicating that the linear JT coupling has a
little influence on the orbital angle in the absence of the
anharmonic effect. This also shows the key role of the anharmonic
effect in the OO from the other aspect.
%
%

\subsection{Magnetic Properties}
   The magnetic properties are also strongly influenced by the
anharmonic effect, as shown in Fig. 5. In the present situation, we
find that the difference between J$_x$ and J$_{y}$ is negligible.
From Fig. 5, one finds that in the absence of the anharmonic effect
(G$=$0), due to the SE coupling, the magnetic couplings between Ag
spins are FM, and nearly remain unchange with the increasing of g.
While for G $<$ 0, due to the SE coupling, the spin coupling between
Ag ions is FM at small g. With the increase of the JT effect, the
magnetic couplings between Ag ions, J$_{x,y}$, vary from FM to AFM,
which is in contradiction with Cs$_{2}$AgF$_{4}$.
For g=4.8 eV/$\AA$ and G=0.75g, as denoted by the dashed vertical
line in Fig. 5, the magnetic couplings in the AgF$_{4}$ plaquette,
{\it i.e.} J$_{x,y}$, are about $-$3.3 meV, very close to the
experimental values, $-$3.793 $\sim$ $-$5.0 meV \cite{NatMat5-561}.
The agreement between the present spin coupling strengths and the
experimental observation is much better than those inferred from the
first-principles electronic structure calculations
\cite{CM18-3281,prb76-054426}.
%
%
%

\begin{figure}[htbp]\centering
\includegraphics[angle=0,width=0.4 \columnwidth]{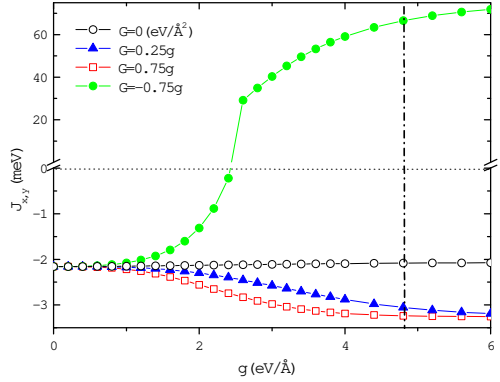}
\caption{(Color online) Magnetic coupling strengthes J$_{x,y}$ in
the AgF$_{4}$ plaquette as a function of the JT coupling with
different anharmonic coupling. The vertical dashed line corresponds
to the case of g=4.8 eV/$\AA$ for Cs$_{2}$AgF$_{4}$. Other
parameters are the same to Fig. 2.}
\label{fig5}
\end{figure}
This result, combining those of the lattice distortion and the orbital angle ,
is in accordance with the Goodenough-Kanamori-Anderson rules \cite{GKA}.
Also in Fig. 5, one notices that in the presence of strong anharmonic coupling
G, the FM coupling strength increases with the linear JT coupling g.
Therefore the FM couplings in Cs$_{2}$AgF$_{4}$ further verify the key role
of the anharmonic effect in the groundstate properties.
In addition, the calculated magnetic moment is about 1 $\mu_{B}$,
which approaches the classical value and is comparable to the
experimental data
observed in muon-spin relaxation experiment \cite{prb75-R220408}.
This is a classical value of Ag ion, in comparison with the
experimental value 0.8 $\mu_{B}$ at 5 K \cite{NatMat5-561}. The
reduction of the experimental magnetic moment may originate from
three reasons: the presence of non-magnetic impurity phases and the
covalence effects as pointed by Mclain, {\it et al.} in Ref.
\cite{NatMat5-561}, as well as the AFM fluctuations between
AgF$_{4}$ layers. We expect that the weak interlayer AFM interaction
and the covalency effect, which are neglected in the present 2D
model, contribute to the major part of the reduction of the magnetic
moment of Ag spins.

\section{REMARKS AND SUMMARY}

   As we have shown in the preceding sections, the anharmonic JT effect
drives the 4d orbitals of Cs$_{2}$AgF$_{4}$ into the
d$_{z^{2}-x^{2}}$/d$_{z^{2}-y^{2}}$ ordering, rather than the
d$_{3z^{2}-r^{2}}$ orbital ordering in La$_{2}$CuO$_{4}$. In
consistent with the Goodenough-Kanamori-Anderson rules \cite{GKA},
the ground state of Cs$_{2}$AgF$_{4}$ is unusual FM insulating with
the orthorhombic structure, different from the {\it Neel} AFM
insulating ground state with the tetrahedral structure in
La$_{2}$CuO$_{4}$. Our results preclude the possibility of the FM
ground state originating from the covalency effect
\cite{prb73-214420}.
One consequence of such a difference is that the electron or hole
doping in La$_{2}$CuO$_{4}$ leads to the strong AFM fluctuations,
which may contribute to the Cooper-pairing glue for the high-T$_c$
superconductivity. We anticipate that the electron doping in
Cs$_{2}$AgF$_{4}$ will lead to weak FM fluctuation, though it will
not contribute to the same SC mechanism as the cuprate
superconductors.

   Notice that in the present intermediate correlated and charge transfer
insulator, the 2D effective SE interactions incorporating the JT
couplings underestimate the covalency effect between Ag 4d and F 2p
orbitals. The covalency effect between Ag and F ions can be well
considered within the first-principles electronic structure
calculations \cite{prb76-024417, prb76-155115, CM18-3281,
prb73-214420}. On the other hand, once the JT distortion occurs, the
hopping integrals, t$_{x,y}$, between Ag ions are slightly different
in different crystallographic axes. We expect that this change will
not significantly modify our results quantitatively.

  In summary, an effective model Hamiltonian with spin, orbital
and lattice degrees of freedom coupling to each other, allows us to
self-consistently describe the lattice structure change, orbital
ordering and magnetic coupling properties, {\it etc.}
Driven by the JT distortion, especially by the anharmonic effect,
Cs$_{2}$AgF$_{4}$ stabilizes in the 2D FM ground state. The correct
lattice structure, orbital ordering and magnetic coupling observed
experimentally can also be addressed in the present theoretical
framework.

\acknowledgements

   This work was supported by the NSFC of China, the BaiRen Project, and
the Knowledge Innovation Program of the Chinese Academy of Sciences.
Part of the calculations were performed at the Center for
Computational Science of CASHIPS and the Shanghai Supercomputer
Center.


\end{document}